\newcommand{\fig}[3]
{\narrowtext
{\begin{figure}[htb]
\epsfxsize=#3cm
\epsfbox{#1}
\caption{#2}
\label{fig:#1}
\end{figure} 
}}

\documentstyle[prb,multicol,aps,epsf]{revtex}
\begin{document}

\title{Oscillations of the superconducting critical current in Nb-Cu-Ni-Cu-Nb junctions}
\author{Y. Blum, A. Tsukernik, M. Karpovski, A. Palevski}
\address{
School of Physics and Astronomy, Tel Aviv University, Tel Aviv 69978, Israel
}
\maketitle
\begin{abstract}
We report on experimental studies of superconductor-ferromagnet
layered structures. Strong oscillations of the critical supercurrent
were observed with the thickness variation of the
ferromagnet. Using known microscopic parameters of Ni, we found reasonable agreement 
between the period of oscillations and the decay of the measured
critical current, and theoretical calculations.
\end{abstract}
\draft
\begin{multicols}{2}
The interplay between superconductivity and ferromagnetism is an
old subject which was studied extensively over decades
\cite{Bulaevskii,Buzdin82,Buzdin92}. The most striking effect in such
systems is the formation of
the so called $\pi$ phase junction in a
superconductor-ferromagnet-superconductor (SFS) structure \cite{Bulaevskii}.
One of the manifestations of the $\pi$ phase is
a non-monotonic variation of the critical temperature \cite{Buzdin92,Radovic}, $T_c$,
with the variation of the ferromagnetic layer
thickness, $d$.  
Oscillatory dependence of $T_c$ vs. $d$ of SF multilayers has been observed
by few groups \cite{Jiang,Muhge,Obi,Lazar}, however, other groups
\cite{Strunk,Aarts,Verbanck} reported monotonic behavior in similar structures.
Interpretations of these experiments suggested the $\pi$ junction
mechanism as well as some other origins of the effect
\cite{Muhge,Aarts,Sangjun}.
The recent observation \cite{Ryazanov2001} of non-monotonic
behavior of the critical current {\it as a function of
temperature} in  weak-ferromagnetic layer of Cu$_x$Ni$_{1-x}$
between two Nb layers is considered as an unambiguous proof of the $\pi$
phase formation.
Another interesting theoretical
prediction \cite{Buzdin82,Buzdin92,Heikkila,Efetov} concerns non-monotonic behavior of the
critical current as a function of the thickness of the ferromagnetic
layer $d$.  According
to the above predictions, the critical current $I_c$ is expected to oscillate
and decay as $d$ is increased.
To the best of our knowledge, such a behavior has not been reported
so far.

In this paper, we present the experimental evidence of
oscillatory behavior of the critical current vs. thickness variation
of ferromagnetic Ni layer. We also show a reasonable agreement between
our data and the
theories \cite{Buzdin82,Efetov} in the appropriate limit 
of $E_{ex}>>\hbar /\tau>>k_BT_c$.
Here, $\tau$ and $E_{ex}$ are the electron relaxation time
and the exchange energy of the ferromagnet, and $T_c$ is the
critical temperature of the superconductor.

We have studied temperature and thickness dependence of the critical
current in Nb-Cu-Ni-Cu-Nb junctions. Additionally, we have studied the
thickness dependence of critical current of similar junctions without
a Ni layer. 
The junctions with $10$x$10\mu$m$^2$
area were fabricated with the standard photolithography
technique. The process contained three stages of lithography: liftoff
of the bottom Nb-Cu layer,
liftoff of the variable thickness Ni or Cu layers, and liftoff of the
top Cu-Nb layer. Nb
films were sputtered using a magnetron gun and {\it in situ} covered
with the Cu layer by thermal evaporation, for preventing the Nb
oxidation. The ferromagnet
layers of Ni were e-gun evaporated in a separate vacuum chamber, and
subsequently covered {\it in situ} by Cu. It is important to emphasize that
all samples were prepared simultaneously. The variation of Ni
thickness for Nb-Cu-Ni-Cu-Nb junction and variation of Cu for Nb-Cu-Nb
junctions was obtained by a specially designed shutter, which
exposed the samples in sequence, so that every sample was
exposed to the evaporating Ni or Cu for additional fragments of time. This
method guaranteed that all the interfaces between each layer in our
multilayer structure are identical, and the only difference between
the samples is their Ni or Cu thickness. The thickness of each Nb layer was
2000\AA. The total thickness of the Cu in the Nb-Cu-Ni-Cu-Nb junctions
was 2400\AA~and the Ni thickness
varied from 10\AA~to 90\AA. 

In the other set of junctions, namely
Nb-Cu-Nb, the Cu thickness varied from 5000\AA~to 10000\AA. The structure of these junctions is shown
schematically in the inset of Fig.\ref{fig:dVdI.eps}. 

Typical resistance of both types of junctions was around
$100\mu\Omega$, independent of both Ni and Cu thicknesses. The
critical temperature of the Nb was about 8.5K.
The measured resistivity of Ni layer $\rho=31 \mu \Omega cm$, together with the known value\cite{Fierz} $\rho l=1.5 \cdot 10^{-11}
\Omega cm^2$, allowed us to
establish the mean free path $l=48$\AA.

\fig{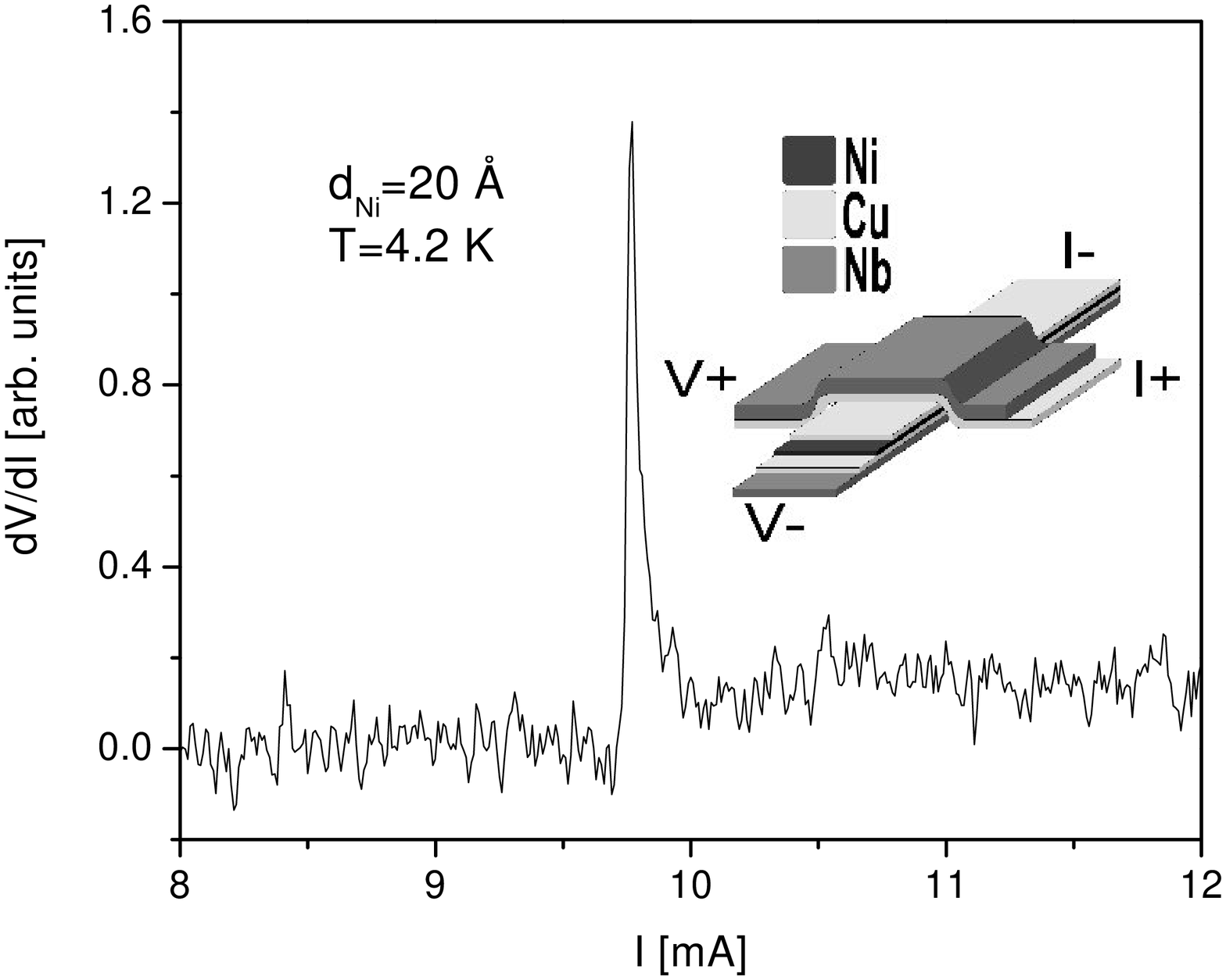}{dV/dI of one of the samples. The critical current was
  defined as the onset of the peak. In the inset: schematic picture of a
  Nb-Cu-Ni-Cu-Nb junction.}{9.}

The measurements were performed in $^4$He cryostat in the range from
6.5K down to
1.5K. The critical current was measured by passing a DC current with a
small AC modulation through the sample. The AC voltage which appeared
above the critical DC current was picked up by lock-in
amplifier operated in a transformer mode.

A typical differential resistance of both types of junctions vs. DC current is
plotted in Fig.\ref{fig:dVdI.eps}, which clearly indicates on the sharp
onset of dissipation at $I=I_c$. 
First, we show results of the Nb-Cu-Nb set (Fig.\ref{fig:Cu.eps}). 
As expected \cite{Zaikin}, the critical
current $I_c$ normalized to the length of the junction $L$, decreases
exponentially with $L$. From the decay of the critical current we
found the thermal
length of the the Cu, $L_T=\sqrt{\hbar D/2\pi k_B T}=1600$\AA~at 4.2K,
and the diffusion constant $D=860cm^2/s$.

\fig{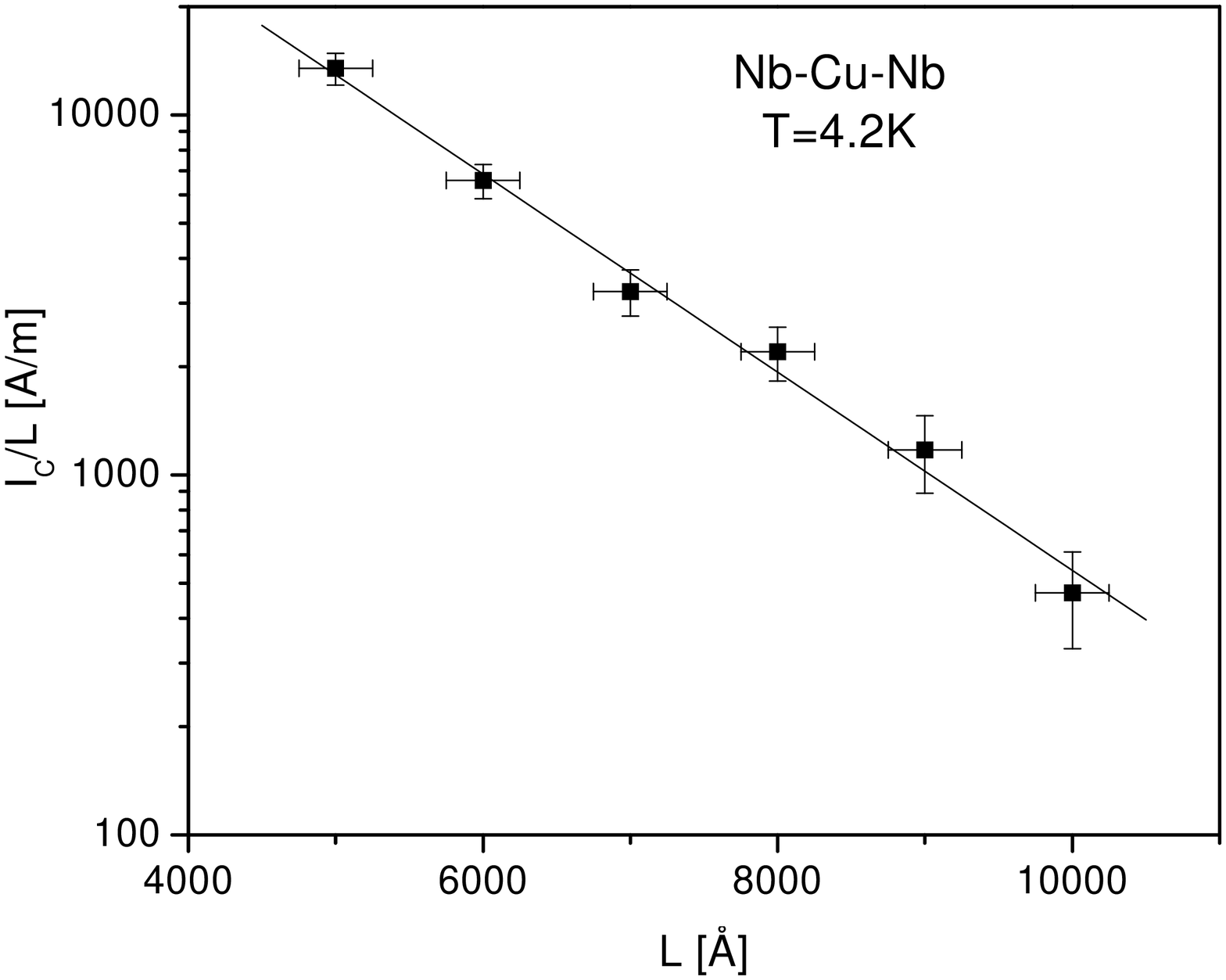}{The normalized critical current $I_c/L$ of a Nb-Cu-Nb
  junctions as a function of the Cu thickness $L$. The straight line 
  represents the theoretical decay.}{9.}

Very different thickness dependence was found in the Nb-Cu-Ni-Cu-Nb
junctions. Fig.\ref{fig:Ic42.eps} shows the thickness dependence of
the critical current in these junctions at $T=4.2K$. In spite of the
large error bars, the non-monotonic variation of the critical
current is quite evident, namely, the deviations of the data from the
exponential decay surmounts by far the uncertainty of each
measured point. 

A further, and even a stronger evidence for the oscillatory behavior is
provided by temperature dependence of the critical
current. Fig.\ref{fig:IcT.eps} shows a family of $I_c$ vs. $T$ curves
for all Ni thicknesses, which are normalized to their values at
$4.2K$. We define the slope $\alpha_T$ of temperature variation, namely
$\alpha_T \equiv d(ln I_c(T))/dT$, and plot these values as a function of $d$ in
Fig.\ref{fig:Ic42.eps} (squares). Oscillations of $\alpha_T$ are very prominent,
and are in anti-phase with the oscillations of the critical
current. Unlike the critical current, which had quite large
experimental error bars, the slope of the temperature
dependence had error bars of only few percents. Such a behavior of
$\alpha_T$ as a function of $d$ is intimately related to the variation
of the critical current oscillations amplitude with
temperature. 
Although the amplitude of the oscillations decreases at high
temperatures, their relative value increases, and on a
semi-logarithmic plot of Fig.\ref{fig:OsciT.eps}, the oscillatory
behavior of $I_c$ is more pronounced at higher temperatures.

\fig{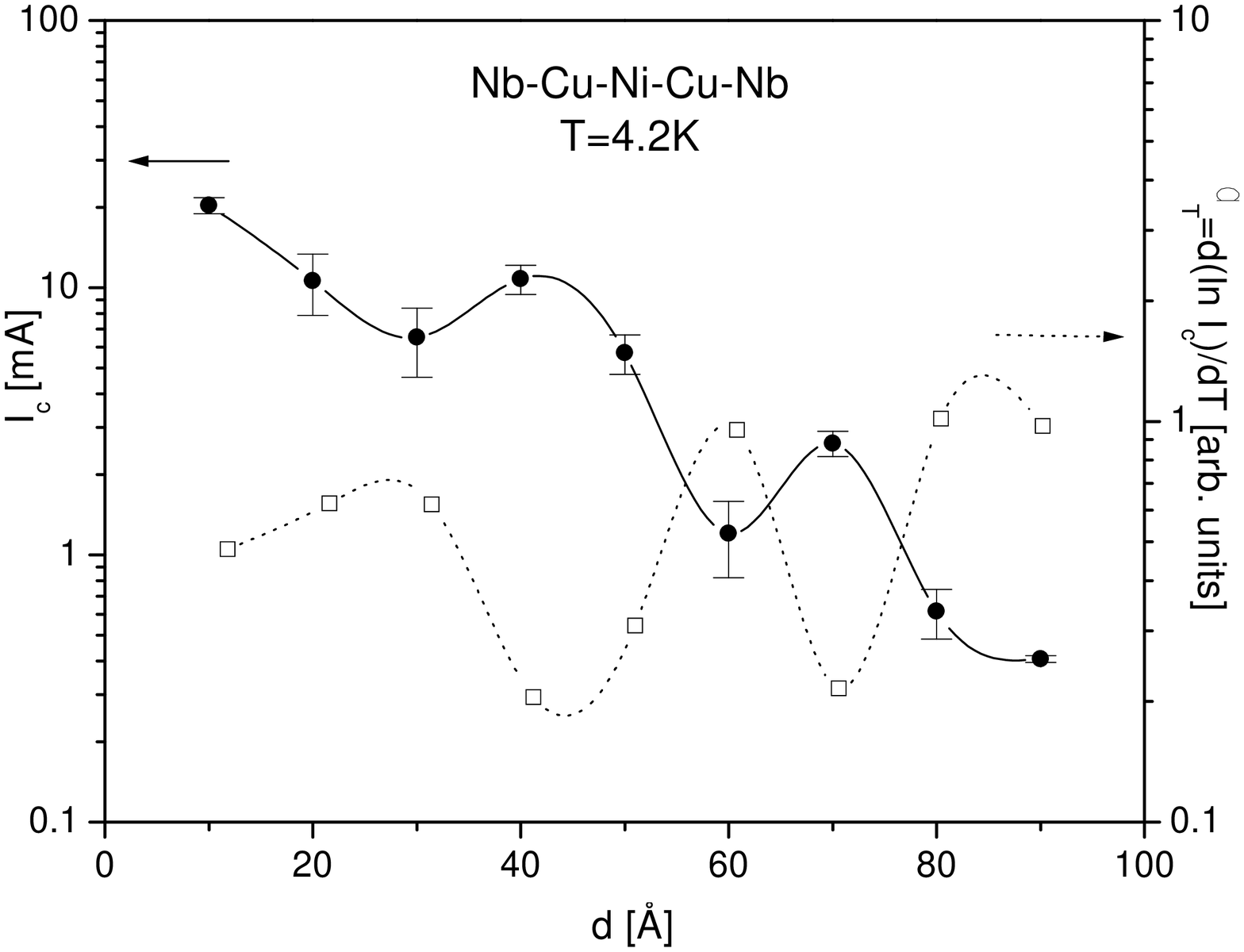}{Critical current of the
  Nb-Cu-Ni-Cu-Nb junctions as a function of the Ni layer's thickness $d$
  at 4.2K (circles). The dependence of the slope $\alpha_T$ on $d$ is
  represented by squares. Both the dashed and dotted lines are only
  for guiding the eye.}{9.}

We would like to start our discussion with what is well known for
superconductor-normal-superconductor (SNS) junctions. The dependence
of critical current of such junctions on the thickness of N layer is
exponential, $I_c~ \alpha~ e^{-L/L_T}$ for $L>L_T$\cite{Zaikin}.
In the opposite limit, where $L<L_T$, $I_c$
approaches a constant value. This behavior has been observed in few
experiments for different material systems
\cite{French2,French}, including our sample presented in
Fig.\ref{fig:Cu.eps}. 

\fig{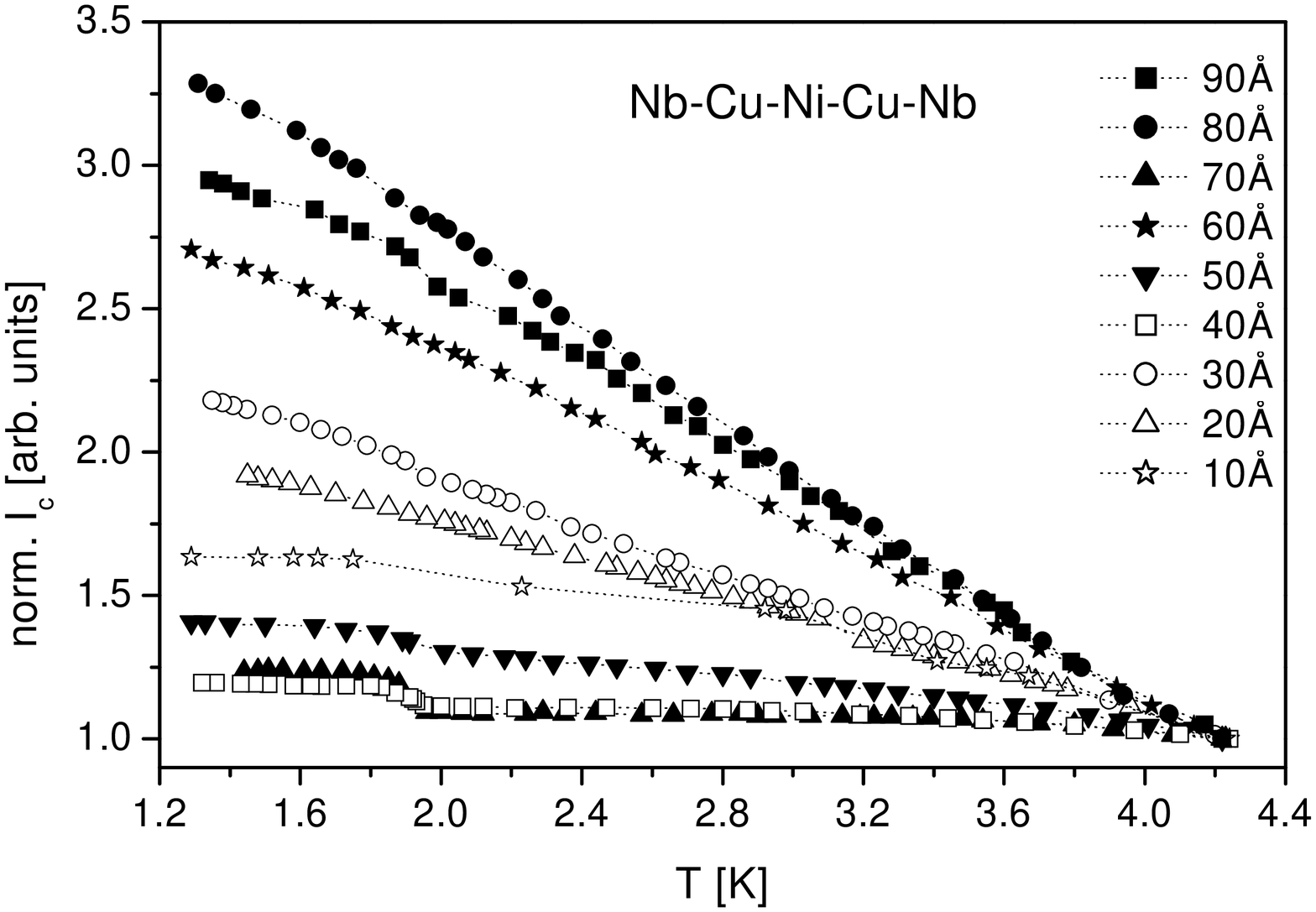}{Critical current as a function of temperature of the
  Nb-Cu-Ni-Cu-Nb junctions for different thicknesses of the Ni layer.}{9.}

Non-monotonic  behavior of the critical current is not expected
theoretically, and had never been observed experimentally in SNS
structures. Since the thickness of Ni is the only
parameter which is varied in our Nb-Cu-Ni-Cu-Nb samples, it is reasonable to assume that the
oscillations observed in the data are due to the presence of Ni 
in the structure. 

As mentioned in the introduction, oscillatory behavior of the critical
current vs. the thickness of the ferromagnetic layer is predicted
theoretically \cite{Buzdin82,Buzdin92,Efetov}. The origin of these oscillations
is the phase shift acquired by electron-hole Andreev particles upon entrance
into the ferromagnet, due to their different spin orientations.
Several expressions have been derived for the critical current in
various limits of the strength of $E_{ex}$, thickness of the
ferromagnet $d$ and disorder. Since in our experiment we have
determined only the magnitude of the critical current $I_c$, the formulae
below will be written for the absolute value $|I_c|$.
For the clean, thin and strong ferromagnetic layers $l>\hbar v_f/E_{ex},d$, the critical
current should vary with the thickness as\cite{Buzdin82}:
\begin{equation}
I_c\sim \left|\sin(2E_{ex}d/\hbar v_f\right)|/(2E_{ex}d/\hbar v_f)
\label{eBuzdinO}
\end{equation}
The dependence of absolute value of the critical current is shown on
Fig.\ref{fig:fits.eps} by dashed line. In the opposite limit, namely
$l<d$, the following dependence was
derived\cite{Buzdin92} (dotted line in Fig.\ref{fig:fits.eps}):
\begin{equation}
I_c \sim y e^{-y}\left|\sin(y+\pi/4\right)| \qquad y=\sqrt{\hbar D/4 E_{ex}}
\label{eBuzdinN}
\end{equation}
Another expression for the critical current in the limit
$l>\hbar v_f/E_{ex}$, is given by\cite{Efetov}:
\begin{equation}
I_c \sim \left| Re \sum_{\omega_n>0}\frac{\Delta^2}{\Delta^2+\omega_n^2}
\int_{-1}^{1} \frac{\mu d\mu}{\sinh(k_{\omega}d/\mu l)}\right|
\label{eEfetov}
\end{equation}
where $\omega_n=\pi T k_B(2n+1)$ is the Matsubara frequency, $n$ is an
integer number,
$k_{\omega}=(1+2|\omega_n|\tau/\hbar)-2iE_{ex}\tau/\hbar$, $\mu=cos
\theta$, $\theta$ is the angle between the momentum and the normal to
the SF interface, and $\Delta$
is the order parameter in the superconductor.
The solid line of Fig. \ref{fig:fits.eps} represents the above expression.
In fitting our data to the expressions for all limits, we have
used\cite{Petrovykh}
$v_f=2.8$x$10^5m/sec$ and $l=48$\AA~(based on the
resistivity measurements of Ni). Therefore, the only fitting parameter
apart from the numerical prefactor was the strength of the exchange
interaction $E_{ex}$. 

The periodicity of oscillations in Eq. \ref{eBuzdinO}, $L_{osc}\sim \pi \hbar v_f/E_{ex} \simeq
54$\AA~fits the best our data when $E_{ex}=107 \pm3 meV$. Note, that $L_{osc}$ is twice
larger than the periodicity observed in Fig.\ref{fig:fits.eps}, due to
the absolute value taken for $I_c$. In order to have similar
periodicity in Eq. \ref{eBuzdinN}, we have used weaker exchange
field $E_{ex}=86meV$. All values are close to the recently reported
value \cite{Petrovykh} $E_{ex}=115meV$.

\fig{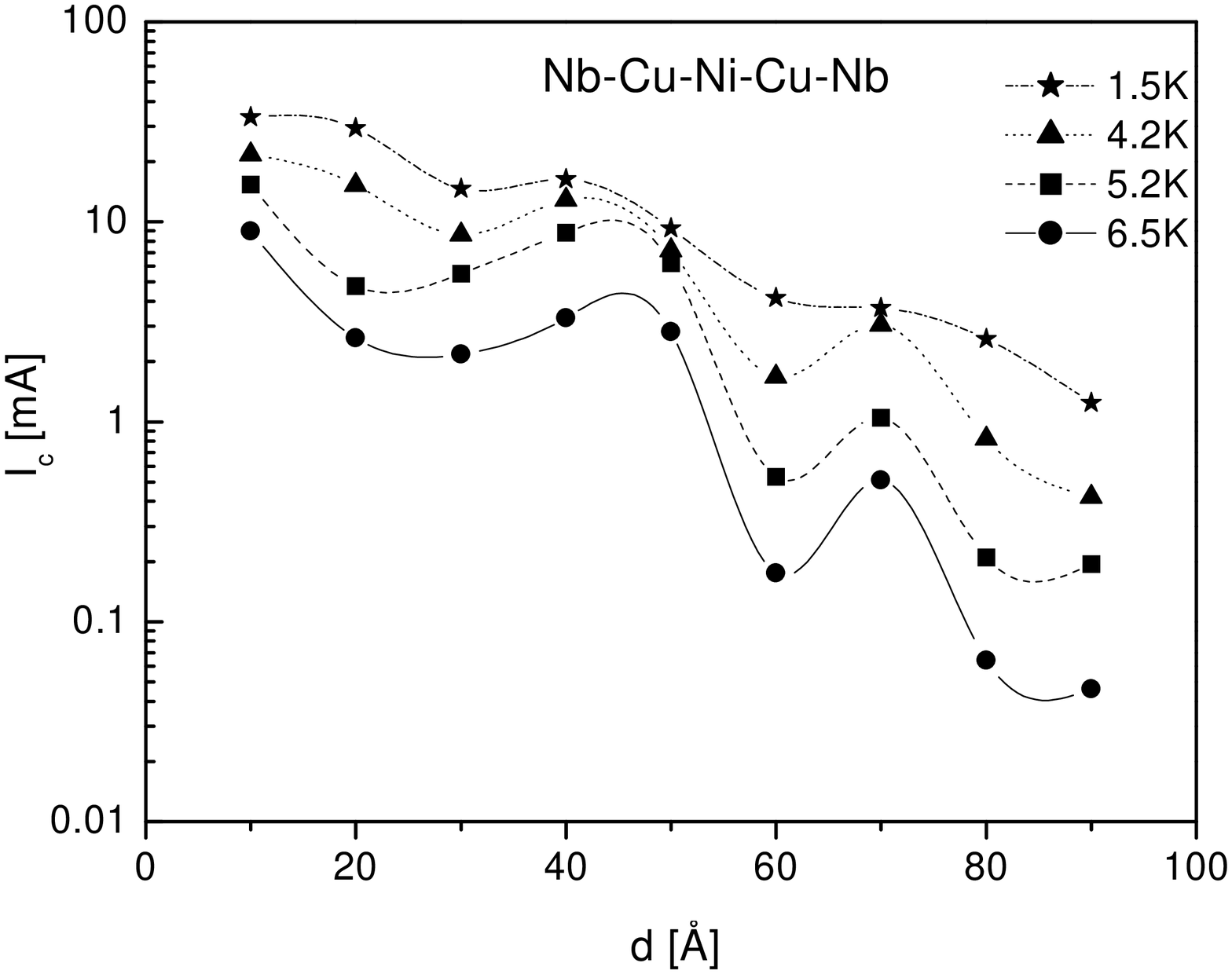}{Critical current as a function of the thickness of the
  Ni layer at different temperatures.}{9.}

The curve given by Eq. \ref{eEfetov} in
Fig.\ref{fig:fits.eps} follows the decay and the phase of the
oscillations of the measured $I_c$ closer than
the other two curves for large $d$.
Furthermore, we estimate $\hbar v_f/E_{ex}=17$\AA~which is smaller
than the measured mean free path $l=48$\AA, and therefore, the
validity of using Eq. \ref{eEfetov}, $l>\hbar v_f/E_{ex}$, is
justified.
However, since Eq. \ref{eEfetov} is valid only for $d>l$, the data
points should follow Eq. \ref{eBuzdinO} for $d<40$\AA.
We thus conclude that our data is consistent with Eq. \ref{eEfetov}
and  Eq. \ref{eBuzdinO} in their appropriate limits.
Therefor, we give the fit of Eq. \ref{eBuzdinO} and Eq. \ref{eEfetov} to our data for four
temperatures in Fig. \ref{fig:fitT.eps}.

\fig{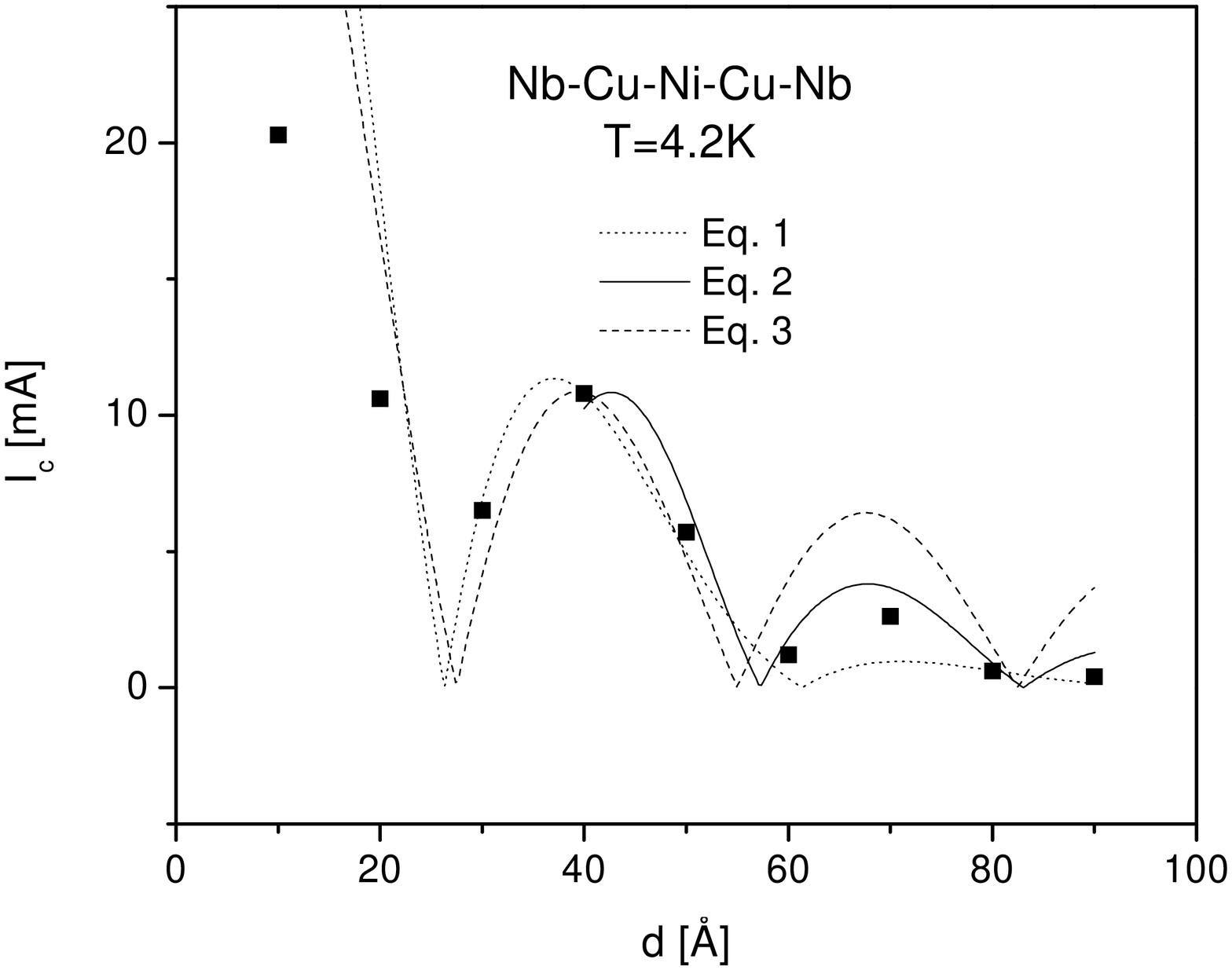}{Critical current of the Nb-Cu-Ni-Cu-Nb junctions at
  4.2K, and the theoretical fits. The dashed line represents
  Eq. 1, the dotted line Eq. 2, and the solid line Eq. 3.}{9.}

\fig{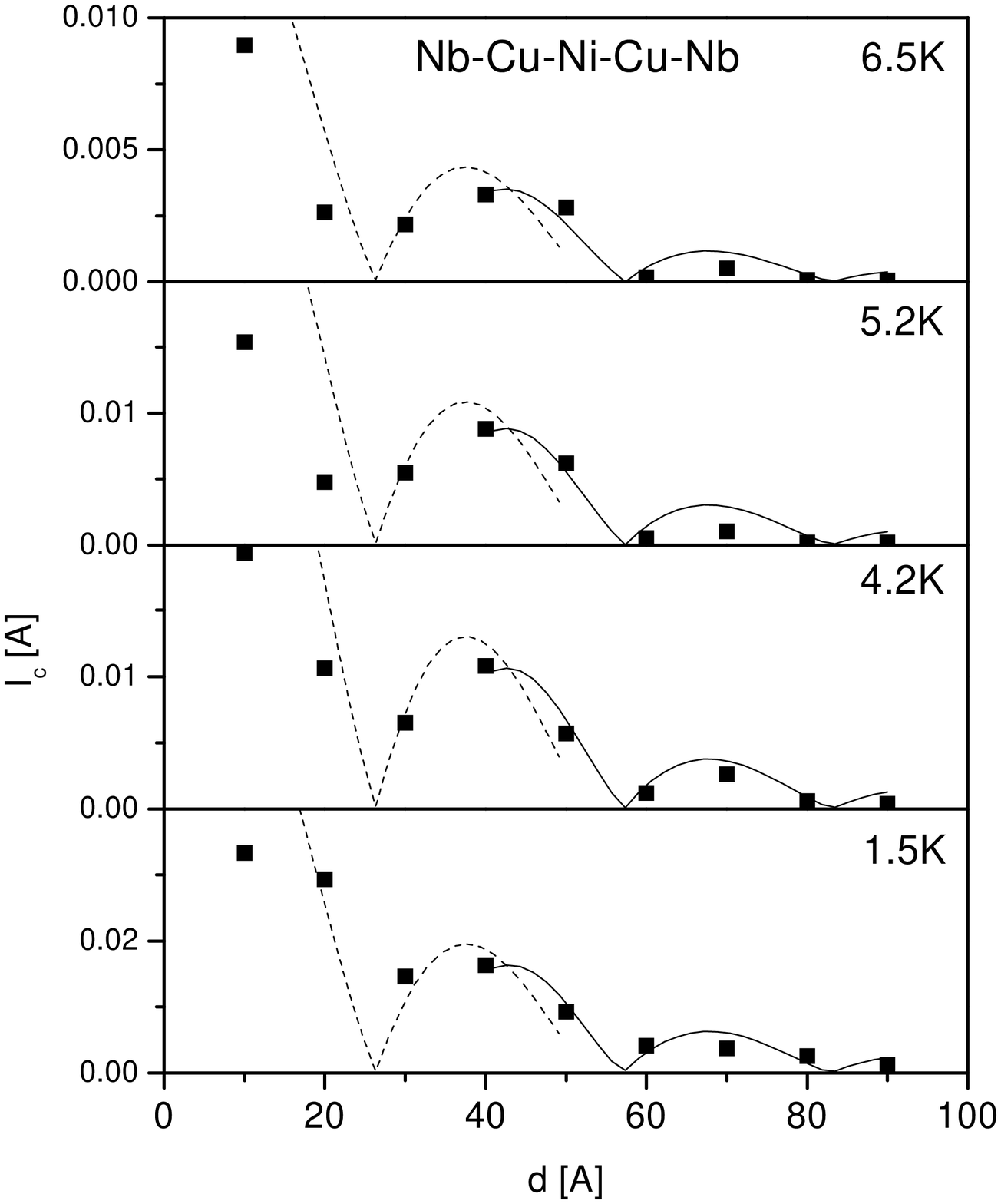}{Critical current of the Nb-Cu-Ni-Cu-Nb junctions at
  1.5K, 4.2K, 5.2K and 6.5K, and the theoretical fits according to
  Eq. 3 (solid line) and Eq.1 (dashed line).}{9.}

The oscillations of $\alpha_T$ (Fig. \ref{fig:Ic42.eps}), which
slightly smear the
oscillations of $I_c$ (Fig. \ref{fig:OsciT.eps}), are not accounted
by any of the mentioned theories and have only a minor effect on the
overall quality of the fit presented in Fig. \ref{fig:fitT.eps}. 
A possible origin of this effect could be related to the presence
of normal metal on both sides of the ferromagnetic layer. The
structure we have used contains layers of Cu of a combined thickness
$L=2400$\AA, which is much larger than the thickness of the Ni layer. It is obvious
that at high temperatures ($L_T<L$) the contribution of Andreev particles
propagating via short diffusion path is predominant, since paths
which are much longer than $L$ do not contribute to the
supercurrent. At low temperatures, where
$L_T>L$, the contribution of longer paths is not
negligible. The latter implies that the impact of the trajectories
crossing Ni layer several times is higher at low temperatures. 
The multiple crossing of the ferromagnetic layer could produce
additional  harmonics of the basic frequency of the oscillations, which will
smear the non-monotonic behavior.
This multiple crossing has not been considered in
the existing theories for S-F-S junctions, and therefore might be responsible for the
observed temperature dependence of the oscillations in S-N-F-N-S
junctions (Fig. \ref{fig:IcT.eps}).

In summary, we have observed the oscillations and the decay of the critical
current in Nb-Cu-Ni-Cu-Nb junctions upon the increase of the thickness
of the Ni layer. We found a reasonable agreement with the recent
theoretical calculations in the appropriate limit.
The puzzling temperature dependence of the oscillations' amplitude remains
unexplained.

We would like to thank A. Aharony, K. B. Efetov, O. Entin-Wohlman,
V. Fleurov, Y. Imry,
K. Kikoin, Z. Ovadiyahu and A. Schiller for fruitful
discussions. Partial support by the Israel Science Foundation founded by the Israel
Academy of Sciences and Humanities --- centers of
Excellence Program is gratefully acknowledged.

\end{multicols}
\end{document}